\documentclass[aps,prd,reprint,superscriptaddress,nofootinbib,
longbibliography,
floatfix
]{revtex4-2}

\usepackage{amsmath}
\usepackage{amssymb}
\usepackage{bm}
\usepackage{graphicx}
\usepackage{float}
\usepackage{dcolumn}
\usepackage{booktabs}
\usepackage{xcolor}
\usepackage{hyperref}

\hypersetup{
    colorlinks=true,
    linkcolor=blue,
    citecolor=blue,
    urlcolor=blue
}

\newcommand{\resultfigure}[2]{%
    \includegraphics[width=0.88\textwidth,trim={#1},clip]{#2}%
}

\begin{document}
\title{Testing selectively enhanced QCD axions couplings as an explanation of the RX J1856.5-3754 hard X-ray excess}

\author{Charul Rathod}
\email{charulrathod1813@gmail.com}
\affiliation{Department of Physics, Birla Institute of Technology and Science - Pilani, Vidya Vihar, Pilani, Rajasthan - 333031, India}

\author{Madhukar Mishra}
\email{madhukar@pilani.bits-pilani.ac.in}
\affiliation{Department of Physics, Birla Institute of Technology and Science - Pilani, Vidya Vihar, Pilani, Rajasthan - 333031, India}

\author{P. K. Das}
\affiliation{Department of Physics, Birla Institute of Technology and Science Pilani, K. K. Birla Goa Campus, Goa - 403726, India}

\date{\today}

\begin{abstract}
We explore the possibility of explaining hard X-ray data obtained from one of the Magnificent Seven (M7) neutron-stars (NSs) employing QCD axion-converted photons. The emission of thermal axions along with neutrinos from the core has been considered. We adopt the nucleon-nucleon bremsstrahlung process as the baseline axion production mechanism and Cooper-pair breaking and formation (PBF) as an additional process. We investigate here whether the selectively enhanced QCD axion coupling model can better explain the hard X-ray excess than the commonly used KSVZ axion model. Our results suggest that meV-mass QCD axions cannot explain the hard X-ray observation within the adopted framework. The enhanced selective model in the micro-eV scale provides closer agreement with the hard X-ray data. We thus conclude that the emission of hard X-rays in the $2-8$ keV range from isolated M7 stars could be explained by QCD axions under an enhanced coupling scenario.

\end{abstract}
\maketitle

\section{Introduction}
\label{sec:introduction}

The QCD axion is one of the best-motivated theoretical extensions of the Standard Model. It is the outcome of the Peccei-Quinn (PQ) mechanism, which dynamically resolves the strong-$CP$ problem~\cite{Peccei:1977hh, Peccei:1977ur}. The axion is also known as a light pseudo-Nambu-Goldstone boson whose mass and possible interaction strengths are calculated primarily by the PQ symmetry-breaking scale $f_a$~\cite{Weinberg:1977ma,PhysRevLett.40.279}. Additionally, QCD axions may be a significant target for laboratory, astrophysical, and cosmological investigations, since it may comprise a portion or all of the cosmological dark matter in the universe~\cite{PhysRevLett.43.103,SHIFMAN1980493,DINE1981199,Zhitnitsky:1980tq, PRESKILL1983127,ABBOTT1983133, DINE1983137,di2020landscape}.\\

The two benchmark theoretical frameworks for "invisible axions" that were developed to address the strong-CP problem in quantum chromodynamics (QCD) are KSVZ~\cite{kim1979weak,shifman1980can} and DFSZ~\cite{dine1981simple,zhitnitsky31possible} models. These models have different nucleon and photon coupling coefficients; the axion mass and its interactions remain correlated by the PQ symmetry-breaking scale $f_a$. This correlation limits the set of coupling coefficients at a fixed axion mass. As a consequence of it, the couplings related to these models cannot be chosen independently at a fixed axion mass. Furthermore, another dark matter candidate very similar to QCD axions called axion-like particles (ALPs) provides a possibility in which its mass and corresponding coupling coefficients pertaining to nucleons and photons can be treated as independent parameters. ALPs are not necessarily associated with the solution of the strong-$CP$ problem~\cite{Marsh:2015xka,Irastorza:2018dyq}. The hard-X-ray excess reported from nearby isolated neutron stars have been explained by using this additional freedom pertaining to ALPs~\cite{Buschmann:2019pfp}.\\

Darm\'e et al.~\cite{darme2021selective} have shown the possibility of an enhancement of the coupling within a QCD-axion framework without violating the PQ symmetry. Their model 'B' preserves the QCD link between \(m_a\) and \(f_a\) by employing a clockwork-inspired chain of \(\lambda+1\) Higgs doublets (heavy scalar states) to produce PQ charges that scale as \(2^\lambda\)~\cite{kaplan2016large,choi2016realizing,giudice2017clockwork,DiLuzio:2020wdo,di2018astrophobic}. The photon coupling is significantly enhanced when the largest-charge doublet is coupled to second-generation quarks, whereas the nucleon couplings get smaller contributions from the
strange and charm-sea contributions. The above projected heavy scalar states arise at energies much smaller than the masses of the particle, resulting in an effective field theory in which the axion interactions are parameterized by the low-energy coefficients \(C_\gamma\), \(C_p\), and \(C_n\)\cite{kim2010axions}. Darm\'e et al. have further shown that selectively enhanced QCD-axion couplings can explain the hard-X-ray interpretation via QCD axions~\cite{darme2021selective}.\\

Neutron stars are considered as the best laboratory for exploring the properties of axionic dark matter because of the highly dense and strongly degenerate nature of its core nuclear matter. The thermal evolution of NSs is determined by the heat capacity of core/crust, neutrino emission, thermal-transport, and surface-photon emission~\cite{doi:10.1146/annurev.astro.42.053102.134013,Page_2004,Page:2005fq}. Additionally, a weakly interacting particle such as axions or ALPs provides additional cooling channels and modify the predicted cooling history~\cite{raffelt1996stars,umeda1998axion}.
Nucleon-nucleon bremsstrahlung and pair breaking and formation (PBF) processes are a possible source of thermal axion emission from the core of NSs~\cite{Hanhart:2000ae,Keller_2013,PhysRevD.93.065044}. The axion-nucleon couplings, the internal temperature, and the superfluid properties of neutron-star matter determine the axion emission rates associated with the above mentioned processes.\\

Axions emitted from the stellar interior may also produce electromagnetic signature. In the strong-magnetosphere, axion-photon mixing can convert emitted axions into X-ray photons~\cite{PhysRevD.37.1237,witte2021axion,Yadav:2024xvf}. The hard-X-ray excess reported from the nearby isolated neutron stars has prompted us to explore an additional scenario involving axion production in the interior of NSs and its subsequent conversion into photons in their magnetosphere~\cite{Dessert_2020,Buschmann:2019pfp}.\\

Yoneyama et al. first reported a keV-energy excess from RX J1856.5$-$3754 M7 NSs~\cite{yoneyama2017discovery}. Dessert et al. subsequently analyzed data from \textit{XMM-Newton} and \textit{Chandra} Telescope and found evidence for a hard-X-ray emission above their dominant thermal spectra~\cite{Dessert_2020}. Buschmann et al. proposed that thermally produced ALPs in NSs interiors converted into photons in their
magnetospheres could also explain this hard X-ray excess~\cite{Buschmann:2019pfp}. Further studies by
Fortin et al. examined hard X-ray data and put constraints on
ALPs from magnetars~\cite{fortin2018constraining,fortin2021magnetars}. Moreover, Witte et al. and Millar
et al. investigated the effects of magnetospheric plasma and strong magnetic fields on axion-photon conversion~\cite{witte2021axion,millar2021axion}. Thus, existing hard X-ray data are primarily analyzed using generic ALPs with independently chosen mass and couplings~\cite{Buschmann:2019pfp,fortin2018constraining}.\\


Although ref.~\cite{darme2021selective} establishes a theoretical mechanism showing the possibility of selectively enhanced QCD axion couplings and found its relevance in explaining hard X-ray observations. However, they did not perform a detailed neutron star cooling calculation including possible emission processes and post-emission processes. The present work is an attempt to address this gap by implementing the enhanced coupling model in NSCool code~\cite{2016ascl.soft09009P} and comparing its predictions with canonical KSVZ and the RX J1856.5$-$3754 hard X-ray excess.\\

In this work, we compare the predictions of the canonical KSVZ axion model with the selective enhanced model (model B) using the same NSs structure and microphysical inputs. The thermal evolution calculation is performed with the NSCool code using the APR equation of state(EoS)~\cite{2016ascl.soft09009P}. Nucleon-nucleon bremsstrahlung and PBF emission processes with density-dependent critical temperature profiles are included for the axion emission.
The resulting redshifted axion spectra are converted into photon spectra in the magnetosphere and compared with the hard-X-ray data from RX J1856.5$-$3754.

This paper is organized as follows. Section~\ref{sec:formalism} presents the
stellar model, axion couplings, emission channels, and conversion prescription. Section~\ref{sec:results} presents the results and discussion of the cooling evolution, converted spectra, and comparison with the observed hard-X-ray excess from RX~J1856.5$-$3754. Finally,
Sec.~\ref{sec:conclusions} summarizes our conclusions and the future outlook of the current work.

\section{Formalism}
\label{sec:formalism}
In the current work, we analyze the hard X-ray data from one of M7 NSs, which is related to our theoretically calculated axion-converted photon flux. The required framework is described here in brief in order to estimate this flux. We begin it by considering a non-rotating $1.4M_{\odot}$ NSs described by an APR EoS~\cite{PhysRevC.58.1804}. The mass, pressure profile, and stellar radius are obtained by solving Tolman–Oppenheimer–Volkoff equations (TOV equations)~\cite{PhysRev.55.364,PhysRev.55.374}. These profiles are then used as input to \texttt{NSCool} to obtain the time evolution of its surface temperature and luminosity of emitted axions~\cite{2016ascl.soft09009P}.\\

This thermal evolution of NSs is described by the following energy-balance equation~\cite{Page_2004,Page:2005fq,2015SSRv..191..239P}:

\begin{equation}
C(T_b^{\infty})\frac{d\ T_b^{\infty}}{dt}
=
-L_\nu^\infty-L_a^\infty-L_\gamma^\infty+H^\infty 
\label{eq:global_cooling}
\end{equation}

where, $T_b^{\infty}=T_b(r)e^{\Phi(r)}$ is the redshifted internal
temperature and $C(T_b^{\infty})$ is the total stellar heat capacity.
$L_\nu^\infty$, $L_a^\infty$, and $L_\gamma^\infty$
denote the redshifted neutrino, axion, and surface-photon luminosities, respectively. $H^\infty$ represents possible internal heating. The present calculation does not include internal heating, therefore $H^\infty=0$.\\

Axion and neutrino luminosities are obtained from the internal temperature $T_b^{\infty}$, whereas surface-photon luminosity is determined by its surface temperature, $T_s^{\infty}$~\cite{2001PhR...354....1Y, Page:2005fq,yadav2024thermal}. $T_s$ and$T_b$ are related and strongly dependent on the composition of heat-blanketing envelope and models used to get their thermal conductivity and opacity. Here, we adopt a non-magnetized iron envelope descibed in ref.~\cite{1983ApJ...272..286G,2015SSRv..191..239P} by Gudmundsson, Pethick, and Epstein, which provides the relation between the internal temperature at the bottom of the envelope ($T_b$) and the effective surface temperature($T_s$) i.e., $T_s=T_s(T_b,g_s)$.\\

The standard neutrino-emission processes remain active throughout the thermal evolution and determine $L_\nu^\infty$. However, the possible axion-production channels are described here to get $L_a^\infty$ in order to solve eq.\ref{eq:global_cooling}.\\

\subsection{QCD Axion Couplings}
\label{subsec:qcd_norm}

For a QCD axion, the mass and coupling coefficients are determined by the
Peccei-Quinn (PQ) symmetry-breaking scale \(f_a\)
~\cite{Peccei:1977hh,Peccei:1977ur,Weinberg:1977ma,PhysRevLett.40.279,
di2016qcd}. The relation is expressed as:
\begin{equation}
m_a
\simeq
6.0\,\mu{\rm eV}
\left(\frac{10^{12}\,{\rm GeV}}{f_a}\right),
\qquad
f_a
\simeq
\frac{6.0\times10^6}{m_a({\rm eV})}\,{\rm GeV},
\label{eq:ma_fa}
\end{equation}
with
\begin{equation}
g_{a\gamma\gamma}
=
C_\gamma\frac{\alpha}{2\pi f_a},
\qquad
g_{aN}
=
C_N\frac{m_N}{f_a}.
\label{eq:qcd_couplings}
\end{equation}

Two QCD axion models; namely KSVZ and DFSZ are very frequently used by researchers. But here we 
use KSVZ and selectively enhanced model 'B' published in~\cite{darme2021selective}.

\subsubsection{\bf KSVZ Model}
In the canonical KSVZ scenario, the PQ symmetry is introduced with a new heavy quark doublet ~\cite{PhysRevLett.43.103,SHIFMAN1980493}. For an electromagnetically neutral heavy quark, $E/N=0$, and the coupling coefficients are $C_p^{\rm KSVZ}=-0.47$, $C_n^{\rm KSVZ}=-0.02$, and $C_\gamma^{\rm KSVZ}=-1.92$~\cite{di2016qcd,DiLuzio:2020wdo}.
The axion production and conversion factors are therefore fixed by the same scale \(f_a\), in contrast to a general ALPs model.

\subsubsection{\bf Selectively Enhanced Model B}

The non-minimal QCD axion benchmark follows model 'B' given in
ref.~\cite{darme2021selective}. It is motivated by clockwork
mechanisms that generate exponentially hierarchical low-energy enhanced couplings from a chain of fields
~\cite{choi2016realizing,kaplan2016large,giudice2017clockwork}. A detailed explanation related to these QCD-axion constructions is discussed in ref.~\cite{DiLuzio:2020wdo}.
This enhancement of coupling coefficients via this model retains the QCD axion $m_a$ and $f_a$ relation, but increases the effective nucleon and photon coefficients through a hierarchy of $PQ$ charges. Under the low-energy approximations implemented here; these coefficients are expressed as:
\begin{align}
C_\gamma^{\rm B} &= 2^\lambda,
\nonumber\\
C_p^{\rm B} &=
2^{\lambda-1}\,0.026\,(1-50\kappa),
\nonumber\\
C_n^{\rm B} &=
2^{\lambda-1}\,0.026\,(1+48\kappa).
\label{eq:modelb_coefficients}
\end{align}
Here \(\lambda\) is the integer that represents the number of enhancement stages in the UV multi-Higgs construction. We took $m_a=18\,\mu{\rm eV}$, $\lambda=15$, and $\kappa=10^{-6}$ for this model.

\subsection{Axion Emission Mechanism}
\label{subsec:source}
Thermal axions are produced in the core of NSs through the nucleon-nucleon bremsstrahlung process and the Cooper pair-breaking formation process (PBF).

\subsubsection{\bf Nucleon-Nucleon Bremsstrahlung Process}

The dominant source of axion emission is nucleon-nucleon bremsstrahlung in the cores of NSs~\cite{PhysRevLett.53.1198,Hanhart:2000ae,Carenza:2019pxu,umeda1998axion,PhysRevD.93.065044,Sedrakian_2019,Buschmann:2021juv}. It is represented by the following interaction process:
\begin{equation} 
N+N\rightarrow N+N+a . 
\label{eq:brem_process} 
\end{equation} 
where, $N$ stands for neutron/proton and $a$ is for axion.
The expression for axion emissivity is written as~\cite{PhysRevLett.53.1198,Yadav:2024xvf,Hanhart:2000ae,Carenza:2019pxu}:
\begin{equation}
\epsilon_{a}^{\mathrm{brem}}
=
\frac{\alpha_a n_B \Gamma_{\sigma} T^3}
     {4\pi m_N^2}
I_{a}^{\mathrm{brem}}.
\label{eq:axion_brem_emissivity}
\end{equation}

Here $I_{a}$ represents an integral given by:
\begin{equation}
I_{a}^{\mathrm{brem}}
=
\int_{0}^{\infty}
s(x)x^2 e^{-x}\,dx,
\label{eq:axion_brem_integral}
\end{equation}
where $\Gamma_{\sigma}$ denotes the nucleon spin-fluctuation rate due
to collisions with other nucleons. In the degenerate limit, it is
given by~\cite{PhysRevLett.53.1198,Yadav:2024xvf}:
\begin{equation}
\Gamma_{\sigma}
=
\frac{4\alpha_{\pi}^{\,2}p_F T^3}
     {3\pi n_B},
\label{eq:spin_fluctuation_rate}
\end{equation}

with
\begin{equation}
s(x)
=
\frac{\left(x^2+4\pi^2\right)|x|}
     {4\pi^2\left(1-e^{-|x|}\right)}.
\label{eq:spin_structure_function}
\end{equation}
Here, $p_F$ is Fermi momentum of the nucleon.

\subsubsection{\bf PBF Process}

When the local internal temperature of material in the core falls below a density-dependent critical temperature $T_{c}(\rho)$, core matter turns into neutron superfluid or proton superconducting phase due to nucleon Cooper pair formation~\cite{Flowers:1976ux,Yakovlev_2005,Page:2005fq}. This continuous formation and breaking of these pairs produces axions, which are described as follows:~\cite{Keller_2013,PhysRevD.93.065044,Sedrakian_2019}:
\begin{equation}
\{NN\}\rightarrow N+N+a,
\qquad
N+N\rightarrow\{NN\}+a.
\end{equation}

We include proton singlet and neutron triplet pairing~\cite{Yakovlev_2005,Page:2005fq} because these are considered as dominant pairing scenarios possible in the core of NSs.

For a proton-singlet pairing, the axion emissivity is expressed as~\cite{Keller_2013,PhysRevD.93.065044}:
\begin{equation}
\epsilon_{a}^{\,{}^1S_0}
=
\frac{8}{3\pi f_a^{2}}\,
\nu_p(0)\,
v_F(p)^{2}\,
T^{5}\,
I_a^{s},
\label{eq:singlet_axion_pbf}
\end{equation}
where $f_a$ is axion decay constant, $T$ is the internal stellar temperature, $\nu_p(0)$ is density of states at the Fermi surface, and $v_F(p)$ is the fermi velocity of proton.

\begin{equation}
\nu_p(0)=\frac{m_p^\ast p_{F,p}}{\pi^2},
\end{equation}
where $m_p^\ast$ and $p_{F,p}$ are the effective mass and Fermi momentum of the proton, respectively.

The integral $I_a^{s}$ is given by~\cite{Keller_2013,PhysRevD.93.065044,Yadav:2024xvf}:
\begin{equation}
    I_a^{s}=
z_n^{5}
\left(
\int_{1}^{\infty}
\frac{y^{3}}{\sqrt{y^{2}-1}}\,
\left[f_F(z_n y)\right]^{2}
\,dy
\right)
\end{equation}

where the dimensionless gap parameter and Fermi-Dirac distribution function are expressed as:

\begin{equation}
z_p=\frac{\Delta_p(T,\rho)}{T},
\qquad
f_F(x)=\frac{1}{e^x+1},
\end{equation}
where $\Delta_p(T,\rho)$ is the density and temperature-dependent
proton ${}^1S_0$ pairing gap and $x = \frac{\omega}{2T}$.

For the neutron triplet pairing, the expression for axion emissivity is~\cite{Keller_2013,PhysRevD.93.065044,Sedrakian_2019}:
\begin{equation}
\epsilon_{a}^{\,{}^3P_2}
=
\frac{2C_n^2}{3\pi f_a^2}\,
\nu_n(0)\,
T^5\,
I_{a}^{p},
\label{eq:triplet_axion_pbf}
\end{equation}
where
\begin{equation}
\nu_n(0)=\frac{m_n^\ast p_{F,n}}{\pi^2}
\end{equation}
is the neutron density of states at Fermi surface. Here,
$m_n^\ast$, $p_{F,n}$, $C_n$, and $g_{an}=C_nm_n/f_a$ denote the
neutron effective mass, Fermi momentum, model-dependent coefficient,
and physical axion-neutron coupling, respectively.\\

The triplet-state control integral is
\begin{equation}
I_{a}^{p}
=
\int \frac{\mathrm{d}\Omega}{4\pi}\,
z_n^{5}(\theta)
\int_{1}^{\infty}
\frac{y^{3}\,\mathrm{d}y}{\sqrt{y^{2}-1}}\,
\left[
f_F\!\left(z_n(\theta)y\right)
\right]^{2}.
\label{eq:triplet_axion_integral}
\end{equation}

Here
\begin{equation}
z_n(\theta)
=
\frac{\Delta_n^{\,{}^3P_2}(T,\rho,\theta)}{T},
\qquad
f_F(x)=\frac{1}{e^x+1}.
\end{equation}

\subsection{Axion Spectrum and Converted Photon Flux}
\label{subsec:axion_spectrum_conversion}
The axion spectrum/or corresponding energy differential emissivity is written as:

\begin{equation}
J_a=\frac{d\epsilon_a}{d\omega}
\label{eq:total_axion_spectrum}
\end{equation}
Here, $\omega=E_\infty$ denotes the axion energy measured by a distant observer.

For a degenerate nucleon-nucleon bremsstrahlung process, the axion spectrum is written as~\cite{Yadav:2024xvf}:
\begin{equation}
J_a^{\mathrm{brem}}(\omega)
\equiv
\frac{d\epsilon_a^{\mathrm{brem}}}{d\omega}
=
\frac{\epsilon_a^{\mathrm{brem}}}{120}\,
\frac{\omega^5}{T^6}
\exp\left(-\frac{\omega}{T}\right).
\label{eq:brem_spectrum}
\end{equation}

For the ${\,p\,{}^{1}S_{0}}$ PBF process, the corresponding spectrum is ~\cite{Keller_2013,PhysRevD.93.065044,Yadav:2024xvf}
\begin{equation}
J_a^{\,p\,{}^{1}S_{0}}(\omega)
\equiv
\frac{d\epsilon_a^{\,p\,{}^{1}S_{0}}}{d\omega}
=
\frac{\mathcal{N}_{p\,{}^{1}S_{0}}}{2\Delta_p}\,
\frac{y^3}{\sqrt{y^2-1}}\,
\left[
f_F\left(\frac{\omega}{2T}\right)
\right]^2,
\label{eq:p1s0_pbf_spectrum}
\end{equation}
where,
$y=\frac{\omega}{2\Delta}$.

It is worthwhile to mention here that, in our work, we determine the energy differential luminosity (called axion flux) by multiplying the corresponding energy differential emissivity to the volume of NSs. It is then used in the next section to get the axion-converted photon flux which is equivalent to the observed hard X-ray flux from one of M7 NSs.  

\subsubsection {\bf Conversion Probability}

Thermal axions produced in the neutron-star core escape the star and
propagate through its magnetosphere, where the transverse magnetic
field induces axion-photon mixing through the interaction,
$\mathcal{L}_{a\gamma}=-(g_{a\gamma}/4)
aF_{\mu\nu}\widetilde F^{\mu\nu}$
~\cite{raffelt1988mixing,Buschmann:2019pfp}. The expression for the approximate
non-resonant conversion probability used in
refs.~\cite{Buschmann:2019pfp,Yadav:2024xvf} is:
\begin{align}
P_{a\rightarrow\gamma}
\simeq{}&
1.5\times10^{-4}
\left(
\frac{g_{a\gamma}}
     {10^{-11}\,{\rm GeV}^{-1}}
\right)^2
\left(
\frac{1\,{\rm keV}}{\omega}
\right)^{0.8}
\nonumber\\
&\times
\left(
\frac{B_0}{10^{13}\,{\rm G}}
\right)^{0.4}
\left(
\frac{R}{10\,{\rm km}}
\right)^{1.2}
(\sin\theta)^{0.4},
\label{eq:axion_photon_probability}
\end{align}
where $g_{a\gamma}=C_\gamma\alpha/(2\pi f_a)$ is the axion--photon
coupling, $\omega$ is the axion energy measured by a distant
observer from the earth, $B_0$ is the polar surface magnetic field, $R$ is the
stellar radius of the NS, and $\theta$ is the angle between the direction of axion propagation
and the magnetic axis. 

\subsubsection{\bf Axion-Converted Photon Flux}
The axion converted photon flux received at a distance $d$ is then given by: 
\begin{equation}
\frac{dF_\gamma}{d\omega}
=
\frac{P_{a\rightarrow\gamma}}
     {4\pi d^2}J_a
\label{eq:converted_photon_flux}
\end{equation}

Equation~\eqref{eq:axion_photon_probability} is valid under light-axion power-law approximation. Resonant, adiabatic and separate finite-mass conversion effects are not included in the current calculation.\\

We evaluated the axion-converted photon flux for one of the M7 stars, RX J1856.5$-$3754, whose $t_{\rm RX}=4.2\times10^{5}\,{\rm yr}$, distance $d=0.123\,{\rm kpc}$ and polar magnetic field
$B_0=2.9\times10^{13}\,{\rm Gauss}$. The predicted photon flux is then compared with XMM--Newton
PN+MOS+\textit{Chandra} hard-X-ray bins performed in the
\(2\)-\(4\), \(4\)-\(6\), and \(6\)-\(8~{\rm keV}\) bands
~\cite{Dessert_2020,Buschmann:2019pfp}.


\section{Results and Discussions}
\label{sec:results}

We compare the predictions for canonical KSVZ model at $m_a=16\,\mathrm{meV}$ with the selectively enhanced model-B benchmark $m_a=18\,\mu\mathrm{eV}$ using the APR EoS for the same NS. The microphysical inputs specified in sec.~\ref{sec:formalism} have been used.
Here we attempt to test whether an astrophysical signal attributed to generic ALPs can be explained by a predictive QCD-axion model without losing consistency with the thermal evolution of the NSs. This makes the current work as a combined test of dark-matter couplings, neutron-star
cooling, and an electromagnetic observable from NSs. \\

The numerical values for the coupling parameters used in the current calculations are listed in table~\ref{tab:benchmark-couplings-results}.
\begin{table}[H]
\centering
\caption{ Nucleon couplings are defined by
$g_{aN}=C_Nm_N/f_a$, and $g_{a\gamma}$ is in $\mathrm{GeV}^{-1}$.}
\label{tab:benchmark-couplings-results}
\begin{tabular}{lcc}
\toprule
Parameters & KSVZ Model & Enhanced Model B \\
\midrule
$m_a$ & $16\,\mathrm{meV}$ & $18\,\mu\mathrm{eV}$ \\
$f_a$ [GeV] & $3.875\times10^8$ & $3.444\times10^{11}$ \\
$C_n$ & $-0.020$ & $4.2600\times10^2$ \\
$C_p$ & $-0.470$ & $4.2596\times10^2$ \\
$C_\gamma$ & $-1.92$ & $3.2768\times10^4$ \\
$g_{an}$ & $-4.849\times10^{-11}$ & $1.162\times10^{-9}$ \\
$g_{ap}$ & $-1.138\times10^{-9}$ & $1.160\times10^{-9}$ \\
$g_{a\gamma}$ [GeV$^{-1}$] & $5.755\times10^{-12}$ & $1.105\times10^{-10}$ \\
\bottomrule
\end{tabular}
\end{table}

\subsection{Thermal evolution}

\begin{figure*}[!t]
\centering
\resultfigure{14.4bp 11.0bp 2.9bp 32.6bp}{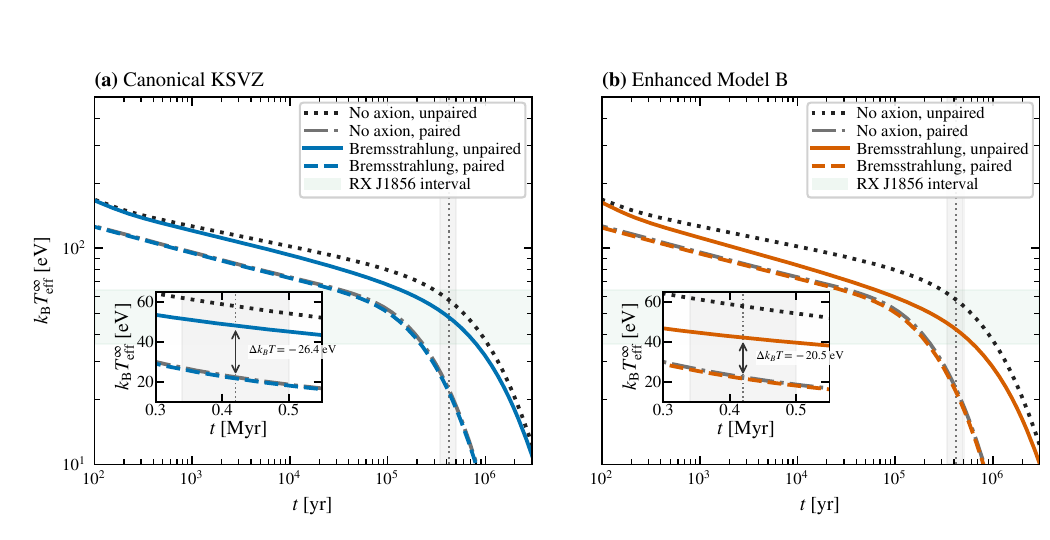}
\caption{Redshifted effective temperature versus age for (a) KSVZ and
(b) enhanced Model B, with and without nucleon pairing for the bremsstrahlung process. The shaded region marks the observed data associated with adopted RX J1856 age and temperature intervals ~\cite{Dessert_2020,Buschmann:2019pfp}.}
\label{fig:teff-brem}
\end{figure*}

\begin{figure*}[!t]
\centering
\resultfigure{14.4bp 11.0bp 2.9bp 32.6bp}{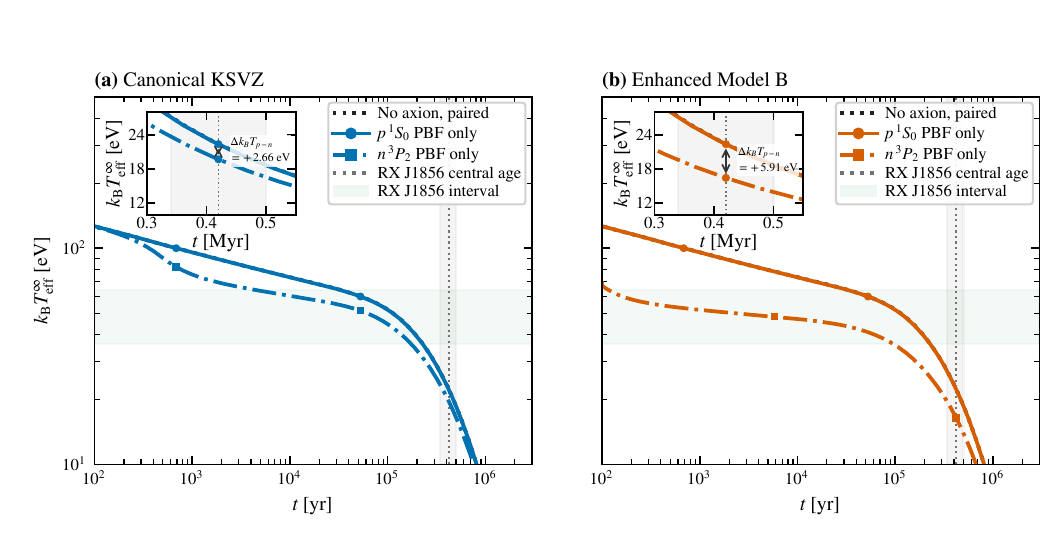}
\caption{Paired redshifted effective temperature versus age for (a) KSVZ and (b) enhanced model 'B'.
Solid and dash-dotted curves show axion via proton ${}^1S_0$ and neutron ${}^3P_2$ PBF processes, respectively. The dotted curve is the no-axion model.}
\label{fig:teff-pbf}
\end{figure*}

Figure~\ref{fig:teff-brem} depicts that the unpaired thermal evolution remains compatible with the observed RX J1856 star temperature interval at the reference age of $0.42\,\mathrm{Myr}$. The no-axion calculation gives $k_{\rm B}T_{\rm eff}^{\infty}=57.86\,\mathrm{eV}$, while axion
bremsstrahlung (unpaired) process gives lower values $48.15\,\mathrm{eV}$ for KSVZ and
$42.17\,\mathrm{eV}$ for enhanced model 'B'. Relative to the results corresponding to the unpaired no-axion case, KSVZ and enhanced model 'B' reduce $k_{\rm B}T_{\rm eff}^{\infty}$ by $16.8\%$ and $27.1\%$, respectively. Thus, enhanced model 'B' shows stronger axion cooling, although both predictions remain within the observed $36$--$64\,\mathrm{eV}$ interval.\\

The bremsstrahlung-paired cooling era leads to a different conclusion. Pairing case suppresses the no-axion prediction to $22.40\,\mathrm{eV}$ while, KSVZ and enhanced model 'B' results to $21.76\,\mathrm{eV}$ and $21.69\,\mathrm{eV}$, respectively. When nucleon pairing is included in bremsstrahlung process, the no-axion, KSVZ, and model 'B' predict nearly the same temperature. Therefore, the cooling in these cases is dominated by pairing, whereas the inclusion of  axions produces only a small additional effect in the cooling era. As a consequence of it, all paired curves fall below the observed RX J1856 thermal interval.\\

The channel-specific thermal evolution due to PBF process shown in Fig.~\ref{fig:teff-pbf} supports the above interpretation of pairing case. Proton $^1S_0$ PBF gives the temperature closer to the paired no-axion model prediction, which is around $22.36\,\mathrm{eV}$ in both axion models. While neutron $^3P_2$ PBF produces a larger accumulated cooling effect, with $19.70\,\mathrm{eV}$ for KSVZ and $16.45\,\mathrm{eV}$ for model 'B'. Thus, lower temperatures obtained for neutron ${}^3P_2$ PBF case gives the strong cooling at earlier ages of the evolution.\\

\subsection{Axion luminosities and Spectra}

\begin{figure*}[!t]
\centering
\resultfigure{8.6bp 11.0bp 48.5bp 32.6bp}{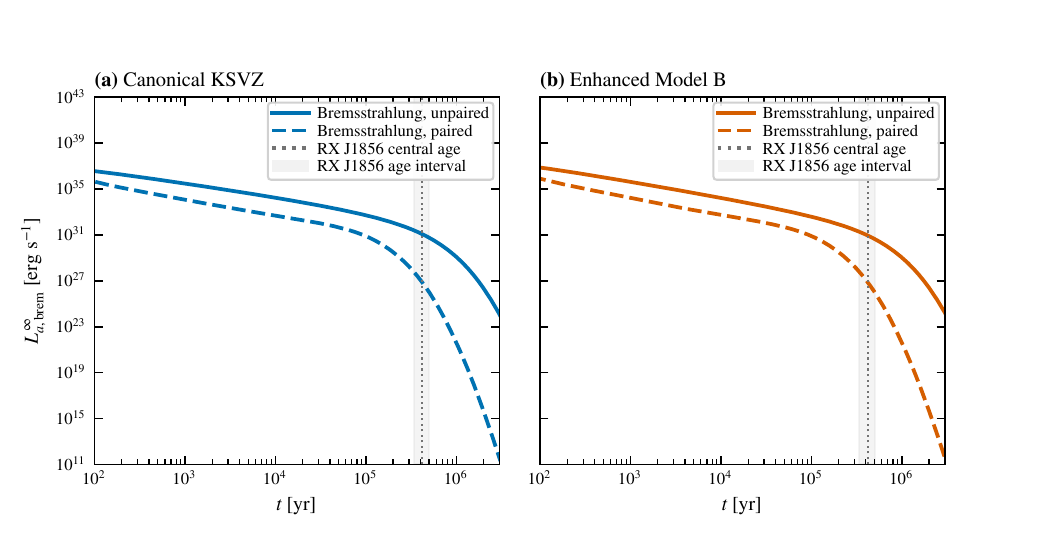}
\caption{Redshifted NN-bremsstrahlung axion luminosity for KSVZ (a) and enhanced model 'B' (b). Solid and dashed curves denote the unpaired and paired cases, while the vertical marker indicates the adopted RX J1856 age.}
\label{fig:lum-brem}
\end{figure*}

The luminosities due to bremsstrahlung with and without nucleon pairing shown in Fig.~\ref{fig:lum-brem} quantify the effect of thermal evolution. At $0.42\,\mathrm{Myr}$, the unpaired gaps for luminosity are
$1.24\times10^{31}\,\mathrm{erg\,s^{-1}}$ for KSVZ and
$9.35\times10^{30}\,\mathrm{erg\,s^{-1}}$ for model B. Despite of enhanced coupling coefficients, model 'B' does not produce a larger, axion luminosity than the corresponding outcome of the KSVZ model. This behavior demonstrates that the enhanced coupling model 'B' luminosity cannot be obtained by simply rescaling the KSVZ result.\\

Nucleon pairing in the bremsstrahlung process reduces the luminosities to $8.09\times10^{26}\,\mathrm{erg\,s^{-1}}$ for KSVZ and
$8.12\times10^{26}\,\mathrm{erg\,s^{-1}}$ for model 'B'. The corresponding
paired-to-unpaired luminosity ratios are $6.51\times10^{-5}$ and
$8.69\times10^{-5}$. Thus, the direct superfluidity effect suppresses bremsstrahlung axion emission, along with producing a colder stellar interior. Therefore, the axion bremsstrahlung luminosity reduces to the pairing case by more than a factor of $10^{4}$ at the age of RX J1856.\\

\begin{figure*}[!t]
\centering
\resultfigure{8.6bp 11.0bp 44.2bp 32.6bp}{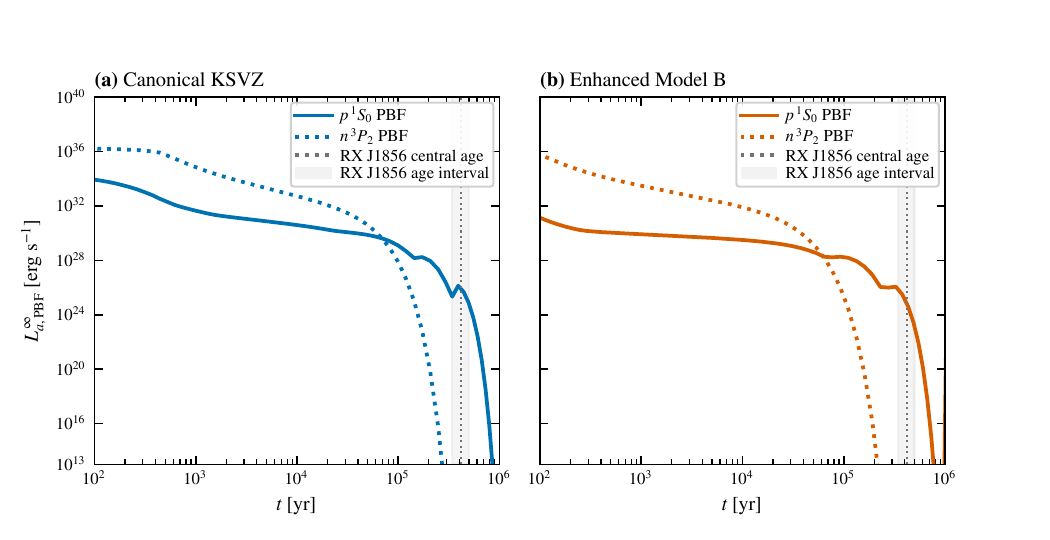}
\caption{Redshifted PBF axion luminosities for KSVZ (a) and enhanced model 'B' (b). Solid and dotted curves denote proton ${}^1S_0$ and neutron ${}^3P_2$ PBF, respectively; the vertical marker indicates the RX J1856 age.}
\label{fig:lum-pbf}
\end{figure*}

Figure~\ref{fig:lum-pbf} shows that proton $^1S_0$ PBF is the only
non-negligible instantaneous PBF process at this age. Its luminosity is
$8.70\times10^{25}\,\mathrm{erg\,s^{-1}}$ for KSVZ and
$1.02\times10^{25}\,\mathrm{erg\,s^{-1}}$ for model 'B', but still lies below the paired bremsstrahlung luminosity in both cases. While the neutron $^3P_2$ component pairing has already switched off to a numerically insignificant value. Its earlier effect can nevertheless account for the colder neutron-triplet cooling in Fig.~\ref{fig:teff-pbf}; the present luminosity and the accumulated cooling effect are different observables.\\

\begin{figure*}[!t]
\centering
\resultfigure{13.9bp 16.3bp 49.0bp 32.6bp}{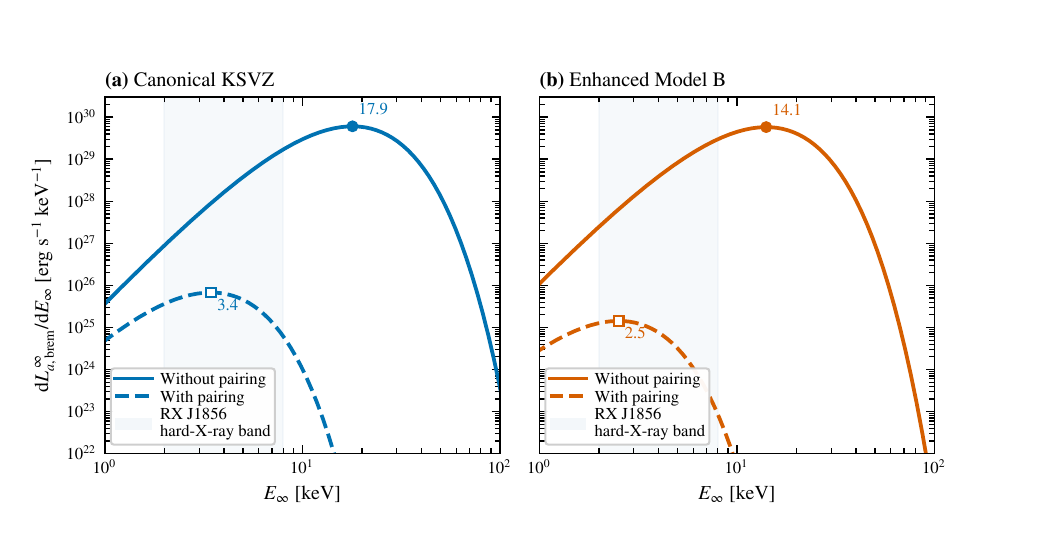}
\caption{NN-bremsstrahlung axion spectra for KSVZ (a)
and enhanced model 'B' (b). Solid and dashed curves denote the unpaired and paired cases, while the shaded region marks the RX J1856 $2-8$ keV band.}
\label{fig:spec-brem}
\end{figure*}

From the axion spectra shown in fig.~\ref{fig:spec-brem}, it is clear that the pairing effect reduces the emission of axion and shifts the spectral peak toward the lower energies.
The unpaired KSVZ and model 'B' spectra peak lie at
$17.9$ and $14.1\,\mathrm{keV}$.
Pairing shifts the peaks to $3.4$ and $2.5\,\mathrm{keV}$, respectively.
But simultaneously reduces the axion flux contained in the observed hard-X-ray energy range.
For KSVZ, in the $2-8$ keV range, the flux decreases from
$3.29\times10^{29}$ to $2.39\times10^{26}\,\mathrm{erg\,s^{-1}}$. For
model 'B' it decreases from $6.45\times10^{29}$ to
$3.24\times10^{25}\,\mathrm{erg\,s^{-1}}$. Pairing results, therefore, have more favorable peak energy, but it may have insufficient intensity to account for the observed hard X-ray flux.\\

\begin{figure*}[!t]
\centering
\resultfigure{16.3bp 19.7bp 2.9bp 31.7bp}{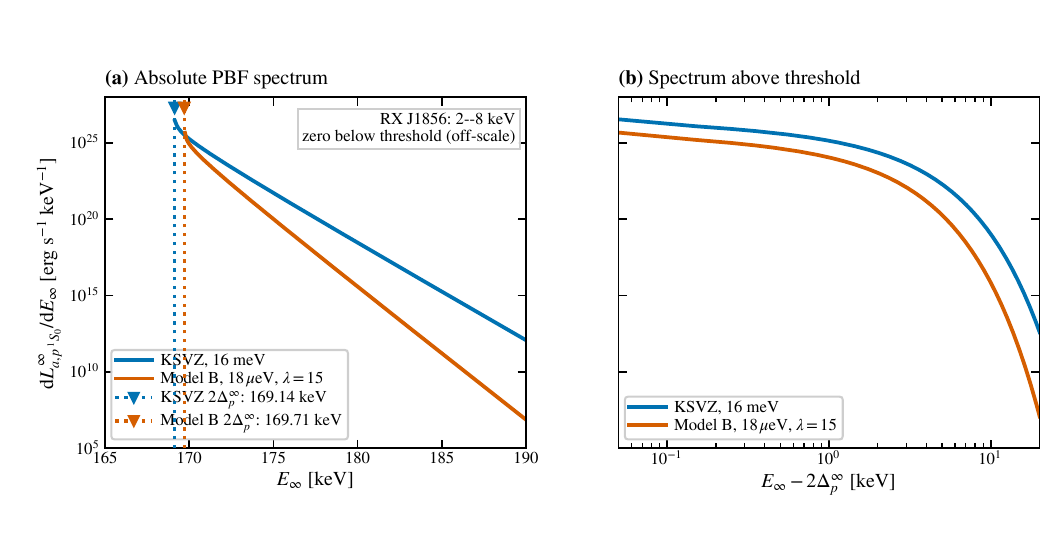}
\caption{Proton ${}^1S_0$ PBF spectra for KSVZ and enhanced model 'B'. Panel (a) shows the redshifted thresholds, while panel (b) displays the spectra above threshold. The RX J1856 2-8 keV band has no support in this
zone.}
\label{fig:spec-pbf}
\end{figure*}

The proton $^1S_0$ PBF axion flux represented in Fig.~\ref{fig:spec-pbf} has redshifted thresholds $169.14\,\mathrm{keV}$ for KSVZ and $169.71\,\mathrm{keV}$ for model 'B'. No flux is observed in the RX J1856 $2-8$ keV band in the adopted core temperature and density-dependent critical temperature profile. It shifts the axion flux and reproduces the corresponding NSCool proton-PBF luminosities to a good numerical precision. Hence, this absence is caused by the selected temperature and
density-dependent gap, not by a normalization failure. Since the calculation performed here is only for one core density and temperature, the absence of PBF emission in this zone does not imply that PBF is absent throughout the entire neutron-star core.\\

\subsection{Converted hard-X-ray flux}

\begin{figure*}[!t]
\centering
\resultfigure{15.4bp 22.6bp 0bp 32.6bp}{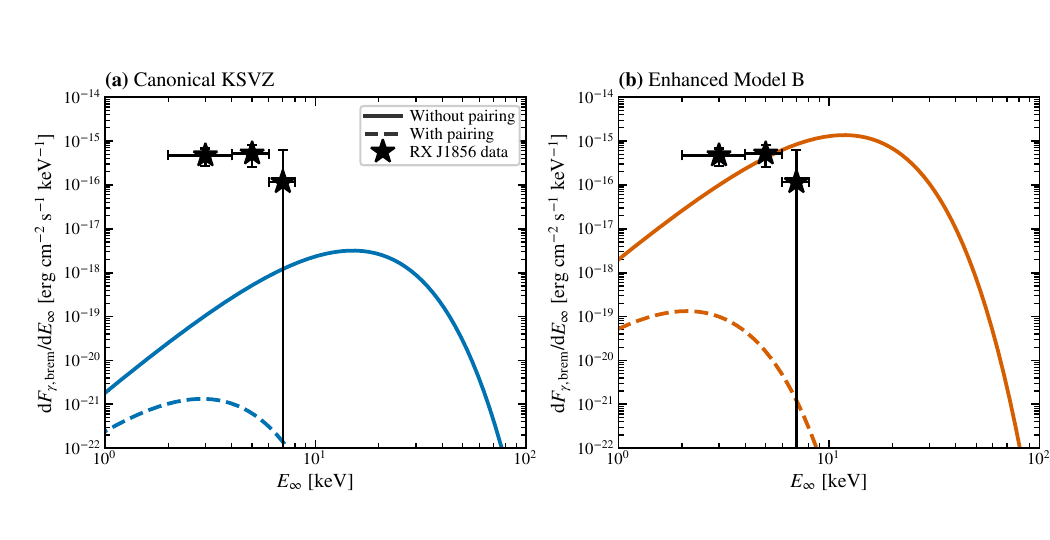}
\caption{Converted bremsstrahlung photon spectra for KSVZ (a) and enhanced Model B (b). Solid and dashed curves denote the unpaired and paired cases, while stars show the RX J1856 data.}
\label{fig:photon-brem}
\end{figure*}

Axion-converted photon flux with respect to energy due to the bremsstrahlung (with and without nucleon pairing) process is shown in fig.~\ref{fig:photon-brem}. For the unpaired case, the calculated photon flux lying in the energy range  $2-8$ keV is
$3.73\times10^{-18}\,\mathrm{erg\,cm^{-2}\,s^{-1}}$ for canonical KSVZ and
$2.74\times10^{-15}\,\mathrm{erg\,cm^{-2}\,s^{-1}}$ for model 'B'. Prediction of selective enhanced model 'B' is seen brighter after conversion from axion to photon. Although its axion luminosity in the same band is only about approximately twice of the value obtained from KSVZ model. The observable enhancement flux is consequently dominated by the larger photon coupling rather than by a comparable increase in axion production. This is an important phenomenological outcome of the selective enhancement coupling model~\cite{darme2021selective}.\\

The paired values remain still below the observed hard X-ray data. The integrated
$2-8$ keV photon fluxes are $4.05\times10^{-21}\,\mathrm{erg\,cm^{-2}\,s^{-1}}$
for KSVZ and $2.35\times10^{-19}\,\mathrm{erg\,cm^{-2}\,s^{-1}}$ for
model 'B'. Hence, for the pairing case that shifts the converted photon flux peaks towards the observed energy range but suppresses the photon signal by approximately three orders of magnitude for KSVZ and four orders of magnitude for model 'B'.\\

\begin{figure*}[!t]
\centering
\resultfigure{23.5bp 23.5bp 1.9bp 32.6bp}{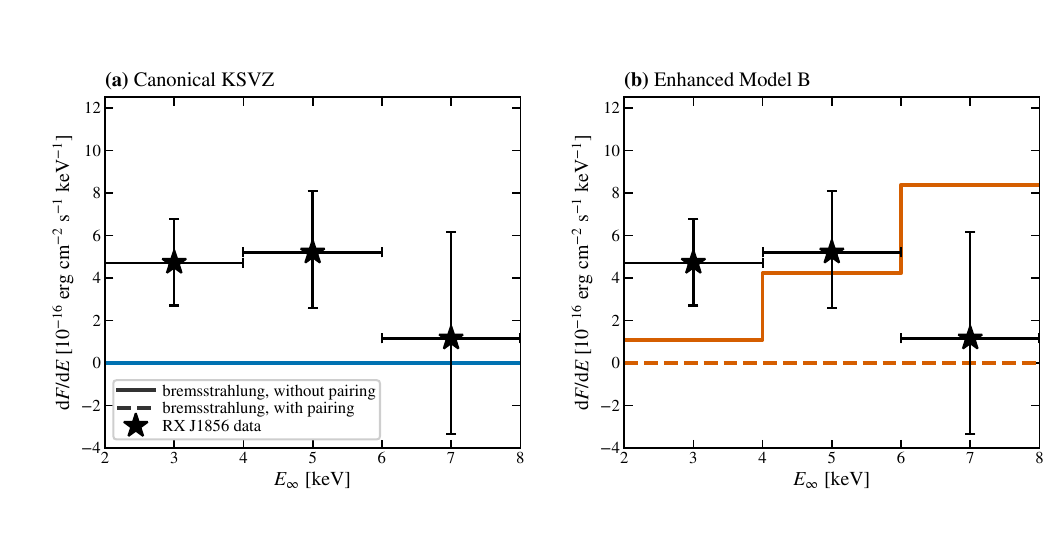}
\caption{Binned RX J1856 hard-X-ray spectrum. Stars show the combined PN,
MOS, and Chandra measurements~\cite{Dessert_2020,Buschmann:2019pfp}; solid
and dashed staircases denote the unpaired and paired bremsstrahlung predictions, respectively.}
\label{fig:stairs-brem}
\end{figure*}

Figure~\ref{fig:stairs-brem} directly compares the model predictions with
the RX J1856.5-3754 hard X-ray excess~\cite{Dessert_2020,Buschmann:2019pfp}. Since the data are reported in the $2-4$, $4-6$, and $6-8$ keV energy bins, the continuous model spectrum is averaged over the same intervals and displayed as a staircase plot. Each step represents the predicted mean differential flux within one observational bin; the jumps between steps are bin boundaries, not physical spectral features. In units of $10^{-16}\,\mathrm{erg\,cm^{-2}\,s^{-1}\,keV^{-1}}$ as shown in plots, the observed central values in the $2-4$, $4-6$, and $6-8$ keV bins are $(4.72,\,5.20,\,1.16)$. For the unpaired KSVZ model the predicted values are $(0.00119,\,0.00528,\,0.0122)$ negligible in every bin. Whereas, the unpaired enhanced model-B prediction, $(1.10,\,4.22,\,8.39)$, reaches the observed normalized data scale and reproduces the middle bin within its quoted uncertainty limits.\\

The photon flux for integrated unpaired model 'B' is only $23.7\%$ above the integral of the three observed central values. This trend shows that selective enhancement removes the total normalization deficit. However, the calculated energy distribution remains inconsistent with the observed spectral shape, particularly in the lower ($2-4$ keV) and higher energy ranges ($6-8$ keV).
A simple analysis based on the displayed asymmetric uncertainties improves for unpaired model 'B', but it is not a substitute for the response-folded in PN, MOS, and Chandra likelihood analysis to explain hard X-ray excess used in the generic-ALPs case investigated by Buschmann et al.~\cite{Buschmann:2019pfp}.\\

\clearpage

\section{Conclusions and Future Outlook}
\label{sec:conclusions}

In this work, we investigate whether fixed QCD-axion benchmarks or a selective enhancement coupling model can account for the hard X-ray excess of RX J1856 while remaining consistent with its thermal evolution. The main findings are summarized as follows:

\begin{enumerate}

\item Within the adopted conversion treatment with and without nucleon pairing for the bremsstrahlung process, the canonical KSVZ model at $m_a=16\,\mathrm{meV}$ produces a hard X-ray flux far below the observed flux data by PN+MOS+Chandra telescope.

\item The unpaired selectively enhanced model 'B' benchmark reaches the
integrated $2-8$ keV flux scale without introducing an additional fitted
normalization. However, it under predicts the $2-4$ keV bin and overpredicts
the $6-8$ keV bin, and therefore does not reproduce the complete spectral
shape.

\item Nucleon pairing shifts the bremsstrahlung spectral maximum toward the observed energy range but suppresses the axion luminosity and converted photon flux by several orders of magnitude. The adopted pairing model also predicts a surface temperature below the RX J1856 thermal interval.

\item Using the adopted density-dependent critical-temperature profile and the local temperature of the selected core zone, the redshifted proton ${}^{1}S_0$ pair-breaking threshold is
$2\Delta_p^\infty\simeq169\,\mathrm{keV}$ for both axion benchmarks.
Because proton PBF emission requires a minimum energy threshold,
$E_\infty\geq2\Delta_p^\infty$, the resulting spectrum does not lie in
the observed $2$-$8\,\mathrm{keV}$ band. Proton ${}^{1}S_0$ PBF, therefore,
does not contribute to the predicted RX J1856 hard X-ray flux in this calculation.

\item Selective enhanced coupling regime can remove the normalization deficit of the canonical QCD-axion prediction, but enhancement alone is insufficient to describe the thermal state and the three-bin hard-X-ray spectrum simultaneously.
\end{enumerate}

These results show reasonably good agreement with data for the selectively enhanced benchmark over canonical KSVZ only at the level of the unpaired hard-X-ray excess. This analysis shows that QCD axions can also explain the hard X-ray data under an enhanced coupling-based model like generic ALPs. Although proton ${}^{1}S_0$ PBF does not contribute to the observed $2$-$8\,\mathrm{keV}$ band, it generates a signal above its redshifted threshold near $169\,\mathrm{keV}$. Future shell-resolved more precise calculations combined with axion-photon conversion and high-energy observational data from NSs may test whether this channel produces a detectable hard-X-ray or soft-$\gamma$-ray signatures. We can also test the sensitivity to the stellar mass, EoS, envelope composition, and pairing models. Finite-mass magnetospheric conversion and a response-folded analysis of the X-ray data will provide a more robust test of the enhanced QCD-axion scenario.\\

\begin{acknowledgments}
The authors thank the Department of Physics, BITS Pilani, Pilani - 333031 (Rajasthan), India, for providing the necessary facilities and administrative support required for the work. One of the authors, Charul Rathod, acknowledges DST New Delhi for providing financial support as a DST INSPIRE fellow.
\end{acknowledgments}

\bibliographystyle{apsrev4-2}
\bibliography{references}
\end{document}